\documentclass[twocolumn,showpacs,amsmath,superscriptaddress,aps]{revtex4-2}

\usepackage{graphicx}
\usepackage{dcolumn}
\usepackage{bm}
\usepackage{hyperref}
\usepackage{siunitx}
\usepackage{color, soul}
\usepackage{lineno}

\begin{document}

\title
{Deterministic Integration of hBN Single-Photon Emitters on SiN Waveguides via Femtosecond Laser Processing}

\author{Daiki~Yamashita}
\email[Corresponding author. ]{daiki.yamashita@aist.go.jp}
\affiliation{Optical Computing Research Team, Photonics-Electronics Integration Research Center, National Institute of Advanced Industrial Science and Technology (AIST), Ibaraki 305-8568, Japan}
\author{Masaki~Yumoto}
\affiliation{Laser Processing Frontier Research Group, Core Manufacturing Technology Research Institute, National Institute of Advanced Industrial Science and Technology (AIST), Ibaraki 305-8568, Japan}
\author{Aiko~Narazaki}
\affiliation{Laser Processing Frontier Research Group, Core Manufacturing Technology Research Institute, National Institute of Advanced Industrial Science and Technology (AIST), Ibaraki 305-8568, Japan}
\author{Makoto~Okano}
\affiliation{Photo-Electronics Integration Research Team, Photonics-Electronics Integration Research Center, National Institute of Advanced Industrial Science and Technology (AIST), Ibaraki 305-8568, Japan}

\begin{abstract}
We demonstrate a post-fabrication method that deterministically integrates hexagonal boron nitride (hBN) single-photon emitters (SPEs) onto silicon nitride (SiN) waveguides. Mechanically exfoliated hBN flakes are dry-transferred onto pre-fabricated SiN waveguides, and localized femtosecond laser irradiation is employed to induce defects with sub-microscale spatial precision. Confocal photoluminescence mapping reveals multiple laser-written bright defects, among which one emitter exhibits narrow spectral linewidth and polarization dependence characteristic of a dipole emitter. The emitter exhibits high brightness and temporal stability, and second-order photon correlation measurements confirm its single-photon nature. Furthermore, we successfully achieve on-chip excitation via the SiN waveguide, demonstrating the compatibility of our approach with mature photonic platform technologies. This deterministic integration technique offers a scalable pathway for incorporating quantum emitters into photonic circuits, paving the way for the development of quantum information processing and communication systems with two-dimensional material hybrid photonic devices.
\end{abstract}
\maketitle

\section{Introduction}

\begin{figure*}[t]
  \includegraphics[width=140mm]{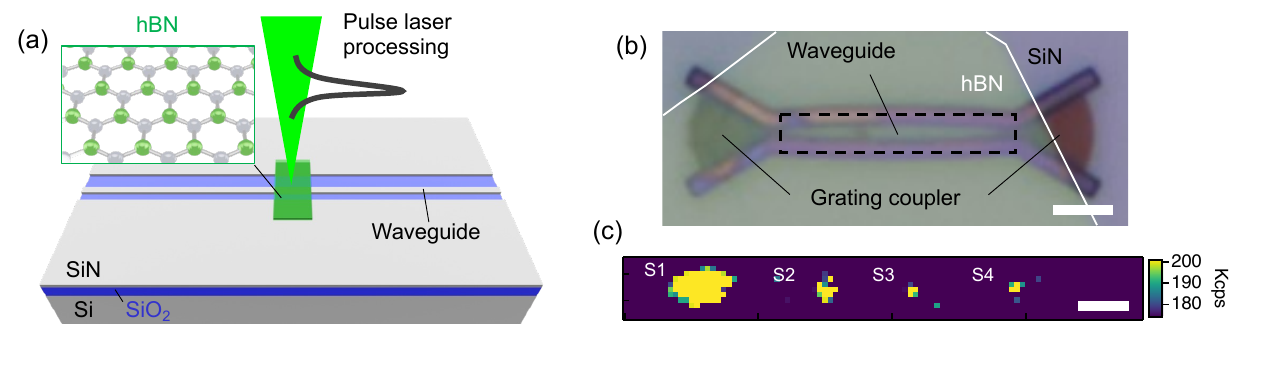}
  \caption{
  \label{Fig1}
  (a) Schematic illustration of the deterministic post-fabrication method based on femtosecond laser processing for integrating hBN SPEs on SiN waveguides.
  (b) Optical microscope image of the device showing an hBN flake transferred onto a SiN waveguide equipped with grating couplers at both ends. The green-colored area indicates the hBN flake. The scale bar corresponds to 5~$\mathrm{\mu}$m.
  (c) Confocal PL map of the region indicated by the dashed box in panel (b) is shown. Four laser-written spots, referred to as S1–S4, are clearly visible. The excitation polarization is set to 90$^{\circ}$ (vertical direction), and the laser power is 2.5~mW. The scale bar is 2~$\mathrm{\mu}$m.
  }
  \end{figure*}

Quantum photonics is rapidly emerging as a foundational technology for next-generation applications, including quantum information processing, quantum communication, and quantum sensing. Among various approaches, photonic chip-based platforms that enable the generation, manipulation, and transmission of quantum states on a single chip have attracted attention due to their scalability and robustness against environmental perturbations~\cite{Wang:2019,Pelucchi:2021,Luo:2023}. The realization of such platforms requires the monolithic integration of essential photonic components such as single-photon sources, waveguides, interferometers, and detectors. In particular, single-photon emitters (SPEs) play a critical role as fundamental elements for quantum communication nodes and photonic qubits in quantum networks~\cite{Uppu:2020,Kim:2020,Elshaari:2020}.

Two-dimensional (2D) materials are promising as host materials for SPEs~\cite{Aharonovich:2022,Turunen:2022}. Among them, hexagonal boron nitride (hBN), known as an ideal insulator with a wide bandgap, enables stable single-photon emission from defect states at room temperature~\cite{Tran:2015,Castelletto:2020,Shaik:2021}. Moreover, optically active defects in hBN can be introduced through various techniques, including thermal annealing~\cite{Li:2019}, chemical treatment~\cite{Mendelson:2019,Lyu:2020}, strain-induced structures~\cite{Proscia:2018,Li:2021}, atomic force microscopy (AFM) indentation~\cite{Xu:2021}, plasma irradiation~\cite{Choi:2016,Xu:2018,Vogl:2018,Gu:2021,Fischer:2021}, foucused ion beam irradiation\cite{Ziegler:2019,Glushkov:2022}, electron beam irradiation~\cite{NgocMyDuong:2018,Fournier:2021}, and laser ablation~\cite{Hou:2017,Gan:2022,Wang:2024}. Furthermore, owing to its van der Waals 2D materials nature, hBN flakes can be transferred onto a wide range of substrates via dry transfer methods, making hBN particularly attractive for applications in integrated photonics~\cite{Su:2024,Kim:2019,Proscia:2020}.

To date, the integration of hBN-based SPEs into photonic chips has primarily involved a strategy in which randomly formed defects are first located by optical characterization, followed by the deterministic fabrication of photonic structures—such as waveguides or cavities—aligned to the defects positions~\cite{Froech:2020,Parto:2022}. Although this method has enabled initial demonstrations of integrated hBN devices, its scalability is limited, and the inherent randomness in defect locations can lead to lower fabrication yields and reduced design flexibility. Consequently, developing deterministic processes that allow precise spatial control over defect formation in hBN remains a critical objective. In response to this need, previous studies have shown that electron beam irradiation can deterministically create SPEs in an hBN-based photonic platform~\cite{Froech:2021,Nonahal:2023,Nonahal:2023a}. However, this platform relies on hBN as both the emitter host and the waveguiding medium, which poses challenges for seamless integration with established photonic technologies.

In this study, we demonstrate a post-fabrication approach for the deterministic integration of SPEs in hBN onto a well-established silicon nitride (SiN) photonic platform. Mechanically exfoliated hBN flakes are dry-transferred onto pre-fabricated SiN waveguides, and optically active defects are subsequently created by localized laser irradiation, enabling precise positioning of SPEs on the waveguides. We show that this method allows {sub-microscale} control over defect locations, improving the practical applicability of quantum emitter integration. The optical characteristics of the fabricated defects are evaluated, confirming single-photon emission. Furthermore, we demonstrate that the SPEs can be excited via the waveguide, enabling on-chip single-photon generation. This approach enables deterministic spatial control over SPE placement and ensures compatibility with well-developed photonic platform technologies, offering a scalable pathway toward the practical implementation of integrated quantum photonic circuits.

\section{Results}

Figure~\ref{Fig1}(a) illustrates a schematic of the deterministic post-fabrication process employed to integrate hBN SPEs onto SiN optical waveguides. SiN is selected as the waveguide material due to its transparency in the visible wavelength range and its natural compatibility with the silicon photonic platform~\cite{Blumenthal:2018,Sharma:2020,Xiang:2022}. Additionally, its refractive index ($\sim2.0$) is close to that of hBN ($\sim2.1$), which minimizes optical coupling losses at the interface. Figure~\ref{Fig1}(b) shows an optical microscope image of the fabricated device, where an hBN flake is transferred onto the central region of a SiN waveguide. The waveguide is designed with a widened center to facilitate laser writing, and grating couplers optimized for a 600~nm wavelength are located at both ends. We utilize mechanically exfoliated hBN flakes and transfer them onto the waveguide using a conventional polydimethylsiloxane (PDMS) stamping technique. Further details of the device fabrication are provided in the Experimental Section.

Defect generation in the hBN flake on the waveguide is carried out by femtosecond laser processing. A pulsed laser (520~nm wavelength, 344~fs pulse width; Spectra Physics: Spirit) is tightly focused onto the sample using an objective lens with a numerical aperture (NA) of 0.95. The laser energy is precisely controlled by rotating a half-wave plate placed before a polarized beam splitter. To form isolated optical defects that host single SPEs, it is crucial to minimize the size of the laser-induced modifications, as larger modification volumes tend to produce multiple emitters~\cite{Hou:2017,Gan:2022,Wang:2024}. To this end, laser processing is carried out near the ablation threshold of materials, where precise and minimal damage can be achieved. We have determined the ablation threshold for the hBN device to be approximately 16.9~nJ, and defect formation is performed using single-shot pulses at this energy. To avoid crosstalk between adjacent emitters, four defects are written along the waveguide with 4~$\mathrm{\mu}$m spacing under the same processing conditions. Additional details on the laser writing process are provided in Section S1 and Fig.~S1 of the Supporting Information.

After laser processing, the hBN flakes are annealed at 900~$^{\circ}$C for 1 hour to optically activate the defects. Figure~\ref{Fig1}(c) presents a confocal photoluminescence (PL) map corresponding to the dashed box in Fig.\ref{Fig1}(b). The sample is excited using a 488~nm continuous-wave (CW) laser focused through an NA = 0.9 objective lens, and the emitted photons are collected using a single-photon avalanche diode (SPAD). The optical measurement setup is detailed in Section S2 and Fig.~S2 of the Supporting Information. Four bright PL spots corresponding to the laser-written locations are clearly observed, confirming that localized emitters can be deterministically formed in hBN on SiN waveguides via femtosecond laser processing.

\begin{figure*}[t]
    \includegraphics[width=160mm]{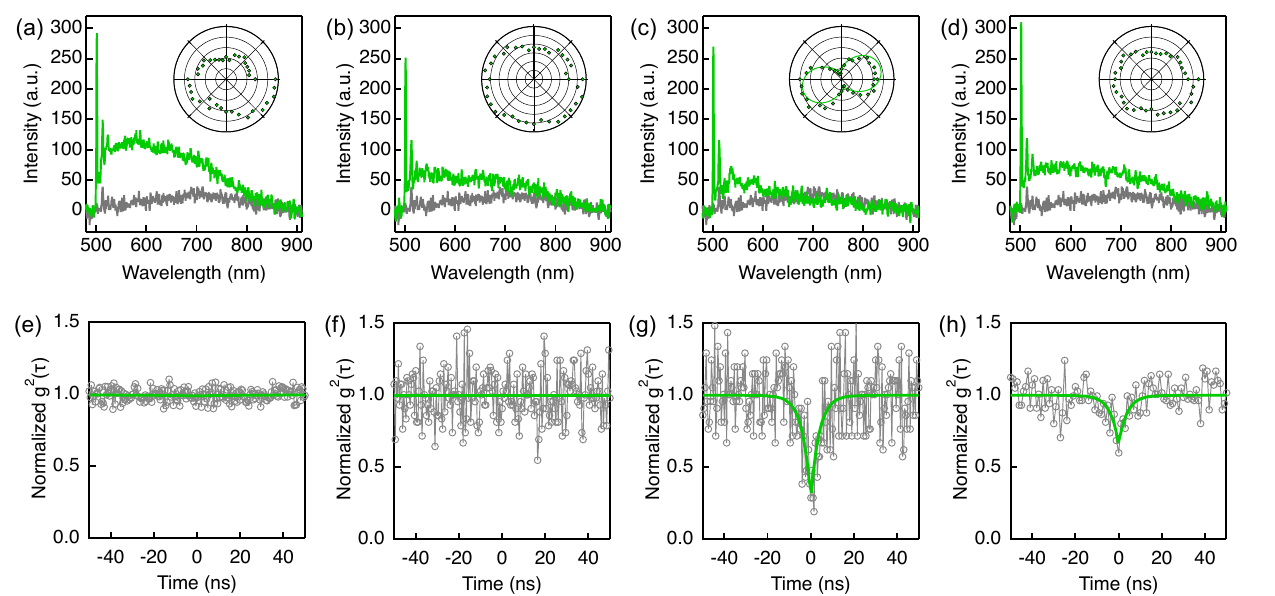}
    \caption{
    \label{Fig2}
    (a–d) PL spectra obtained from S1–S4 defects as indicated in Fig.~\ref{Fig1}(c). In the green curves, the sharp peaks around 500~nm correspond to Raman signals from silicon (500~nm), SiN (512~nm), and hBN (524~nm), while the broad peaks centered at 500–600~nm originate from PL of hBN. The gray curves are background emission from a bare SiN waveguide. For each defect, the excitation polarization angle is adjusted to the value yielding maximum PL intensity, as shown in the respective insets. In the device image of Fig.~\ref{Fig1}(b), the angle of 0$^{\circ}$ corresponds to horizontal polarization. The polarization curve for the S3 defect is fitted using the equation $\cos^2(\theta)$, and the fit yielded an angle $\theta = 16^\circ$. The excitation power for all measurements is 2.5~mW.
    (e–h) Second-order correlation functions $g^{(2)}(\tau)$ mesured for S1–S4 defects. Experimental data are shown in gray, and the fitted curves are in green. The fitting is performed using the equation $g^{(2)}(\tau) = 1 - a e^{-|\tau|/\tau_{1}}$. The excitation power is 2.5~mW for the S1, S2, and S4 defects, and 1.0~mW for the S3 defect. The dependence of $g^{(2)}(\tau)$ on excitation power for the S3 defect is presented in Fig.~S3 of the Supporting Information. Antibunching is observed for the S3 and S4 defects, with $g^{(2)}(0) = 0.28 \pm 0.10$ and $0.66 \pm 0.07$, respectively.
    }
    \end{figure*}

We characterize the optical properties of the four laser-written defects on the SiN waveguide. Figures~\ref{Fig2}(a–d) display the PL spectra from each defect (S1–S4). Raman peaks around 500~nm, and PL emission from hBN centered around 500–600~nm are observed. Gray curves represent the background spectrum from a bare SiN waveguide without hBN. While the S1, S2, and S4 defects exhibit broad PL spectra, the S3 defect shows a narrow spectrum with peaks at 537~nm and 582~nm, corresponding to the zero-phonon line and phonon sideband of hBN, respectively~\cite{Jungwirth:2016,Tawfik:2017}. The insets of Figs.~\ref{Fig2}(a–d) show the polarization dependence of PL intensity under linearly polarized excitation. The S3 defect displays a clear polarization visibility, consistent with dipole emission from a single localized defect, whereas S1, S2, and S4 show weaker polarization visibilities.

To investigate single-photon properties, we perform second-order photon correlation measurements using the Hanbury Brown–Twiss (HBT) setup as shown in Fig.~S2. Figures~\ref{Fig2}(e–h) show the second-order correlation function $g^{(2)}(\tau)$ for the S1–S4 defects. Clear antibunching is observed in Figs.\ref{Fig2}(g) and \ref{Fig2}(h), which correspond to the S3 and S4 defects, respectively. The S3 defect (Fig.\ref{Fig2}(g)) exhibits $g^{(2)}(0) < 0.5$, confirming its nature as a single-photon source. In contrast, the S1, S2 and S4 defects show $g^{(2)}(0) > 0.5$, suggesting emission from multiple emitters.

\begin{figure}[t]
\includegraphics[width=70mm]{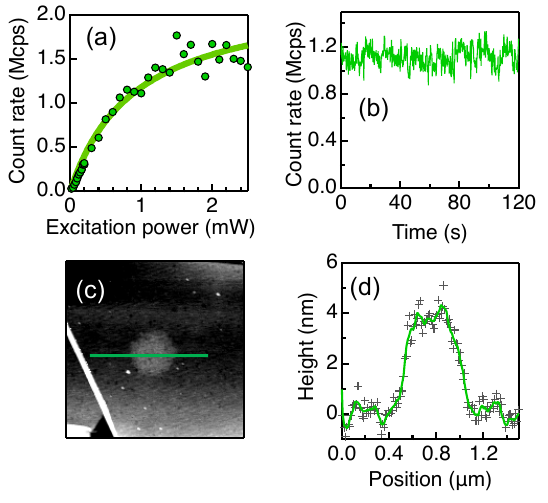}
\caption{
\label{Fig3}
(a) PL count rate from the S3 defect as a function of excitation power. Data points are shown as circles, and the green line represents a fit using the standard saturation model.
(b) Time-trace PL intensity of S3 at 1~mW excitation power, demonstrating stable emission. The time bin width is set to 200~ms.
(c) AFM image of the S3 defect. The green scale bar is 1.5~$\mathrm{\mu}$m.
(d) Height profile along the green line in panel (c). Measured data points are plotted as crosses, and the green curve indicates a box-averaged profile using a 5-point moving window.
}
\end{figure}

For the S3 defect, we further evaluate the emitter performance and morphology. Figure~\ref{Fig3}(a) shows the PL count rate $I$ as a function of excitation power $P$. The data are fitted using the saturation model $I = I_{\mathrm{sat}} \cdot P / (P + P_{\mathrm{sat}})$, yielding a saturation count rate $I_{\mathrm{sat}}$ = 2.26 Mcounts/s and a saturation power $P_{\mathrm{sat}} = 0.92$~mW. This brightness is comparable to previously reported laser-fabricated hBN SPEs~\cite{Gan:2022,Wang:2024} and demonstrates that our laser-writing method is capable of producing high-brightness quantum emitters on SiN waveguides. Figure~\ref{Fig3}(b) shows the temporal stability of the emission at 1~mW excitation power. The intensity remains stable at approximately 1.11 Mcounts/s with 7$\%$ fluctuation, and no blinking or bleaching is observed.

Figure~\ref{Fig3}(c) presents the AFM image of the S3 defect, and Fig.~\ref{Fig3}(d) shows the height profile of the defect. The full-width at half-maximum (FWHM) is 480~nm, and the height is 3.7~nm, indicating a bubble-like morphology. The feature size is smaller than the diffraction-limited spot size $D = 1.22\lambda/\mathrm{NA} \sim 668$~nm, suggesting a contribution from nonlinear effects in the femtosecond laser processing. AFM images of S1, S2, and S4 are shown in Fig.~S4 of the Supporting Information. Among these, S2 also exhibits a bubble-like feature, whereas no clear structure is observed for S1 and S4 defects. The observed morphological variability, despite identical laser energy, is attributed to substrate and hBN flake inhomogeneities and possible misalignment during laser writing~\cite{Hou:2017,Gan:2022,Wang:2024}.

\begin{figure}[t]
\includegraphics[width=85mm]{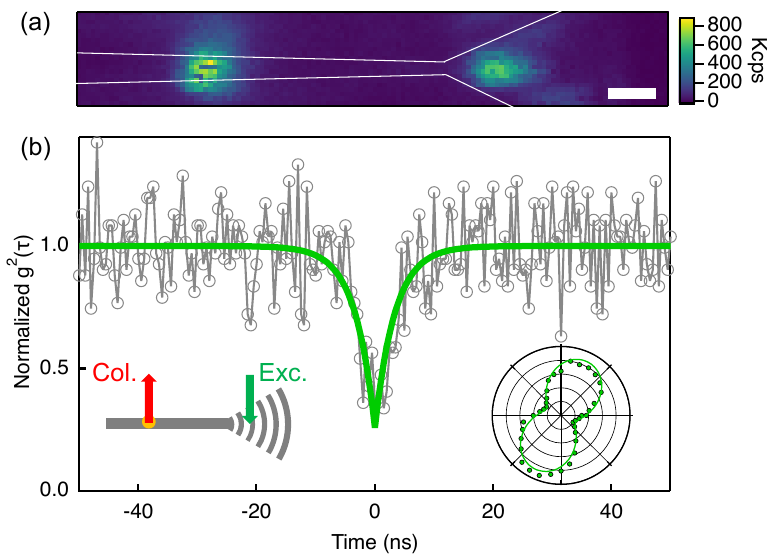}
\caption{
\label{Fig4}
(a) PL intensity map measured by scanning the excitation position using a 4f beam-steering system, while keeping the collection point fixed at the S3 defect. The scale bar is 2~$\mathrm{\mu}$m.
(b) Second-order correlation function $g^{(2)}(\tau)$ of the S3 defect under waveguide excitation. In this configuration, excitation light is coupled into the right-side grating coupler, and emission is collected at the defect spot. Experimental data and the corresponding fit are shown in gray and green, respectively. The data is fitted using the equation $g^{(2)}(\tau) = 1 - a e^{-|\tau|/\tau_{1}}$, yielding $g^{(2)}(0) = 0.26 \pm 0.08$. The excitation power is 2.5~mW.
The lower-left inset shows a schematic diagram of the excitation and collection configuration.
The lower-right inset displays the polarization dependence of the PL intensity under waveguide excitation. Measured data are plotted as circles, and the green line represents a fit using the equation $\cos^2(\theta)$, yielding an angle $\theta = 64^\circ$.
}
\end{figure}

Finally, we demonstrate on-chip excitation of the SPE via waveguide coupling. Figure~\ref{Fig4}(a) presents a PL intensity map measured by scanning the excitation position using a 4f beam-steering system (see Fig.~S2, Supporting Information), while the collection spot is fixed at the S3 defect. In addition to excitation position at S3 defect, a strong PL signal is observed at the right-side grating coupler, indicating successful excitation via the SiN waveguide. Figure~\ref{Fig4}(b) shows the second-order correlation function $g^{(2)}(\tau)$ measured under waveguide excitation. A clear antibunching dip is observed, confirming that the S3 defect retains its single-photon emission characteristics under indirect excitation. The lower-right inset of Fig.~\ref{Fig4}(b) displays the polarization dependence of the PL intensity under waveguide excitation. The maximum excitation angle is shifted compared to that observed under direct excitation (Fig.~\ref{Fig2}c), which we attribute to the grating coupler’s optimal coupling condition occurring at a polarization angle of 90$^\circ$. This result indicates that efficient waveguide excitation requires alignment between the polarization of the emitter and the grating coupler.

\section{Conclusion}
We have demonstrated a deterministic post-fabrication method for integrating SPEs in hBN onto pre-fabricated SiN waveguides. By employing localized femtosecond laser irradiation followed by high-temperature annealing, optically active defects are generated in hBN with sub-microscale spatial control. Optical characterization reveals that one of the laser-written defects exhibits narrow-linewidth emission, strong polarization dependence, and clear photon antibunching, confirming its single-photon nature. The emitter also shows high brightness and temporal stability, indicating its suitability for quantum photonic applications. Furthermore, we have successfully demonstrated on-chip excitation of the hBN SPE via the SiN waveguide, validating the compatibility of the proposed method with mature photonic integration technologies. This waveguide-based excitation preserves the single-photon characteristics of the emitter and highlights the potential for scalable and efficient quantum photonic circuits.

In this study, we focus on exciting a single SPE via a waveguide. Extending this scheme to multiple SPEs integrated along the same waveguide could enable simultaneous multiple single-photon generation. Additionally, while we demonstrate waveguide-based excitation, efficient coupling of the emitted single photons into the waveguide remains a key challenge. Preliminary measurements under collection via the grating coupler reveal an increased contribution from SiN background emission, along with insufficient signal intensity for photon correlation measurements. To improve single-photon collection efficiency, it will be important to align the polarization of the emitter with the waveguide mode or to enhance coupling via optical resonator structures~\cite{Kim:2019,Proscia:2020,Froech:2020,Froech:2021,Parto:2022,Nonahal:2023}. These advancements will expand the capability of hBN-based quantum emitters and support their implementation in integrated quantum photonic systems.

\section*{Experimental Section}
\paragraph*{Sample Preparation.}
Devices comprising SiN waveguides with grating couplers at both ends were fabricated. The waveguides, 20.87~$\mathrm{\mu}$m in length, incorporated an elliptical taper that increases the width from 500~nm at both ends to 1260~nm at the center to facilitate laser processing. The grating couplers were designed for a center wavelength of 600~nm, with a grating period of 438.8~nm, a grating waveguide width of 283~nm, and a total of 12 grating periods. The SiN waveguides were fabricated on silicon wafers with a 3~$\mathrm{\mu}$m thick thermally grown SiO$_2$ layer. A 300~nm thick SiN film was deposited via low-pressure chemical vapor deposition (LPCVD). The waveguide and grating patterns were defined by electron-beam lithography and transferred onto the SiN layer using inductively coupled plasma (ICP) etching with CHF$_3$ gas. hBN crystals were purchased from HQ Graphene. hBN flakes were prepared on PDMS sheets via mechanical exfoliation. Flakes with a thickness of approximately 30~nm were dry-transferred onto the SiN waveguides using a conventional PDMS stamping technique. To remove organic residues, the samples underwent ultraviolet ozone cleaning for 30 minutes followed by annealing at $500~^{\circ}$C for 1 hour.

\paragraph*{Optical Characterization.}
Optical measurements were performed using a homebuilt micro-PL setup as shown in Fig.~S2. A CW 488nm laser was used for excitation, focused onto the sample by a 0.9~NA objective after reflection by a 500~nm-cutoff dichroic mirror. Excitation power and polarization were controlled using a neutral density filter and a half-wave plate, respectively. The sample was placed on a 3-axis feedback-controlled stage for PL mapping. A 4f beam-steering setup enabled independent scanning of a excitation spot.

PL emission passed through the dichroic mirror and a 500~nm long-pass filter, then routed via a flipper mirror to either a multi-mode fiber (MMF) or single-mode fiber (SMF). MMF-coupled light was analyzed with a spectrometer (YIXIST: YSM-8103-04-01), while SMF-coupled light was split by a 50:50 fiber beam splitter and detected by two SPADs (MPD: PDM FC), with photon events recorded using a time-correlated single-photon counting (TCSPC) module (Swabian Instruments: Time Tagger Ultra). The single-photon collection efficiency was 1.75$\%$ (see Section S2, Supporting Information). All measurements were conducted at room temperature under ambient conditions.

\section*{Acknowledgments}
This work wa supported in part by JSPS (KAKENHI JP22K14624, JP24K08296), MEXT (ARIM JPMXP1224AT0098, JPMXP1224UT1040), and Amada Foundation (AF-2024235-C2, AF-2023209-B2).

\section*{Author contributions}
D.Y. carried out sample preparation and performed measurements. M.Y., A.N. and M.O. aided the construction of measurement setup. D.Y. wrote the manuscript with all authors providing input and comments on the manuscript.


\begin{thebibliography}{42}%
    \makeatletter
    \providecommand \@ifxundefined [1]{%
     \@ifx{#1\undefined}
    }%
    \providecommand \@ifnum [1]{%
     \ifnum #1\expandafter \@firstoftwo
     \else \expandafter \@secondoftwo
     \fi
    }%
    \providecommand \@ifx [1]{%
     \ifx #1\expandafter \@firstoftwo
     \else \expandafter \@secondoftwo
     \fi
    }%
    \providecommand \natexlab [1]{#1}%
    \providecommand \enquote  [1]{``#1''}%
    \providecommand \bibnamefont  [1]{#1}%
    \providecommand \bibfnamefont [1]{#1}%
    \providecommand \citenamefont [1]{#1}%
    \providecommand \href@noop [0]{\@secondoftwo}%
    \providecommand \href [0]{\begingroup \@sanitize@url \@href}%
    \providecommand \@href[1]{\@@startlink{#1}\@@href}%
    \providecommand \@@href[1]{\endgroup#1\@@endlink}%
    \providecommand \@sanitize@url [0]{\catcode `\\12\catcode `\$12\catcode `\&12\catcode `\#12\catcode `\^12\catcode `\_12\catcode `\%12\relax}%
    \providecommand \@@startlink[1]{}%
    \providecommand \@@endlink[0]{}%
    \providecommand \url  [0]{\begingroup\@sanitize@url \@url }%
    \providecommand \@url [1]{\endgroup\@href {#1}{\urlprefix }}%
    \providecommand \urlprefix  [0]{URL }%
    \providecommand \Eprint [0]{\href }%
    \providecommand \doibase [0]{https://doi.org/}%
    \providecommand \selectlanguage [0]{\@gobble}%
    \providecommand \bibinfo  [0]{\@secondoftwo}%
    \providecommand \bibfield  [0]{\@secondoftwo}%
    \providecommand \translation [1]{[#1]}%
    \providecommand \BibitemOpen [0]{}%
    \providecommand \bibitemStop [0]{}%
    \providecommand \bibitemNoStop [0]{.\EOS\space}%
    \providecommand \EOS [0]{\spacefactor3000\relax}%
    \providecommand \BibitemShut  [1]{\csname bibitem#1\endcsname}%
    \let\auto@bib@innerbib\@empty
    \bibitem [{\citenamefont {Wang}\ \emph {et~al.}(2019)\citenamefont {Wang}, \citenamefont {Sciarrino}, \citenamefont {Laing},\ and\ \citenamefont {Thompson}}]{Wang:2019}%
      \BibitemOpen
      \bibfield  {author} {\bibinfo {author} {\bibfnamefont {J.}~\bibnamefont {Wang}}, \bibinfo {author} {\bibfnamefont {F.}~\bibnamefont {Sciarrino}}, \bibinfo {author} {\bibfnamefont {A.}~\bibnamefont {Laing}},\ and\ \bibinfo {author} {\bibfnamefont {M.~G.}\ \bibnamefont {Thompson}},\ }\bibfield  {title} {\bibinfo {title} {Integrated photonic quantum technologies},\ }\href {https://doi.org/10.1038/s41566-019-0532-1} {\bibfield  {journal} {\bibinfo  {journal} {Nat. Photonics}\ }\textbf {\bibinfo {volume} {14}},\ \bibinfo {pages} {273} (\bibinfo {year} {2019})}\BibitemShut {NoStop}%
    \bibitem [{\citenamefont {Pelucchi}\ \emph {et~al.}(2021)\citenamefont {Pelucchi}, \citenamefont {Fagas}, \citenamefont {Aharonovich}, \citenamefont {Englund}, \citenamefont {Figueroa}, \citenamefont {Gong}, \citenamefont {Hannes}, \citenamefont {Liu}, \citenamefont {Lu}, \citenamefont {Matsuda}, \citenamefont {Pan}, \citenamefont {Schreck}, \citenamefont {Sciarrino}, \citenamefont {Silberhorn}, \citenamefont {Wang},\ and\ \citenamefont {Jöns}}]{Pelucchi:2021}%
      \BibitemOpen
      \bibfield  {author} {\bibinfo {author} {\bibfnamefont {E.}~\bibnamefont {Pelucchi}}, \bibinfo {author} {\bibfnamefont {G.}~\bibnamefont {Fagas}}, \bibinfo {author} {\bibfnamefont {I.}~\bibnamefont {Aharonovich}}, \bibinfo {author} {\bibfnamefont {D.}~\bibnamefont {Englund}}, \bibinfo {author} {\bibfnamefont {E.}~\bibnamefont {Figueroa}}, \bibinfo {author} {\bibfnamefont {Q.}~\bibnamefont {Gong}}, \bibinfo {author} {\bibfnamefont {H.}~\bibnamefont {Hannes}}, \bibinfo {author} {\bibfnamefont {J.}~\bibnamefont {Liu}}, \bibinfo {author} {\bibfnamefont {C.-Y.}\ \bibnamefont {Lu}}, \bibinfo {author} {\bibfnamefont {N.}~\bibnamefont {Matsuda}}, \bibinfo {author} {\bibfnamefont {J.-W.}\ \bibnamefont {Pan}}, \bibinfo {author} {\bibfnamefont {F.}~\bibnamefont {Schreck}}, \bibinfo {author} {\bibfnamefont {F.}~\bibnamefont {Sciarrino}}, \bibinfo {author} {\bibfnamefont {C.}~\bibnamefont {Silberhorn}}, \bibinfo {author} {\bibfnamefont {J.}~\bibnamefont {Wang}},\ and\ \bibinfo {author} {\bibfnamefont {K.~D.}\ \bibnamefont {Jöns}},\ }\bibfield  {title} {\bibinfo {title} {The potential and global outlook of integrated photonics for quantum technologies},\ }\href {https://doi.org/10.1038/s42254-021-00398-z} {\bibfield  {journal} {\bibinfo  {journal} {Nat. Rev. Phys.}\ }\textbf {\bibinfo {volume} {4}},\ \bibinfo {pages} {194} (\bibinfo {year} {2021})}\BibitemShut {NoStop}%
    \bibitem [{\citenamefont {Luo}\ \emph {et~al.}(2023)\citenamefont {Luo}, \citenamefont {Cao}, \citenamefont {Shi}, \citenamefont {Wan}, \citenamefont {Zhang}, \citenamefont {Li}, \citenamefont {Chen}, \citenamefont {Li}, \citenamefont {Li}, \citenamefont {Wang}, \citenamefont {Sun}, \citenamefont {Karim}, \citenamefont {Cai}, \citenamefont {Kwek},\ and\ \citenamefont {Liu}}]{Luo:2023}%
      \BibitemOpen
      \bibfield  {author} {\bibinfo {author} {\bibfnamefont {W.}~\bibnamefont {Luo}}, \bibinfo {author} {\bibfnamefont {L.}~\bibnamefont {Cao}}, \bibinfo {author} {\bibfnamefont {Y.}~\bibnamefont {Shi}}, \bibinfo {author} {\bibfnamefont {L.}~\bibnamefont {Wan}}, \bibinfo {author} {\bibfnamefont {H.}~\bibnamefont {Zhang}}, \bibinfo {author} {\bibfnamefont {S.}~\bibnamefont {Li}}, \bibinfo {author} {\bibfnamefont {G.}~\bibnamefont {Chen}}, \bibinfo {author} {\bibfnamefont {Y.}~\bibnamefont {Li}}, \bibinfo {author} {\bibfnamefont {S.}~\bibnamefont {Li}}, \bibinfo {author} {\bibfnamefont {Y.}~\bibnamefont {Wang}}, \bibinfo {author} {\bibfnamefont {S.}~\bibnamefont {Sun}}, \bibinfo {author} {\bibfnamefont {M.~F.}\ \bibnamefont {Karim}}, \bibinfo {author} {\bibfnamefont {H.}~\bibnamefont {Cai}}, \bibinfo {author} {\bibfnamefont {L.~C.}\ \bibnamefont {Kwek}},\ and\ \bibinfo {author} {\bibfnamefont {A.~Q.}\ \bibnamefont {Liu}},\ }\bibfield  {title} {\bibinfo {title} {Recent progress in quantum photonic chips for quantum communication and internet},\ }\href {https://doi.org/10.1038/s41377-023-01173-8} {\bibfield  {journal} {\bibinfo  {journal} {Light: Sci. Appl.}\ }\textbf {\bibinfo {volume} {12}},\ \bibinfo {pages} {175} (\bibinfo {year} {2023})}\BibitemShut {NoStop}%
    \bibitem [{\citenamefont {Uppu}\ \emph {et~al.}(2020)\citenamefont {Uppu}, \citenamefont {Pedersen}, \citenamefont {Wang}, \citenamefont {Olesen}, \citenamefont {Papon}, \citenamefont {Zhou}, \citenamefont {Midolo}, \citenamefont {Scholz}, \citenamefont {Wieck}, \citenamefont {Ludwig},\ and\ \citenamefont {Lodahl}}]{Uppu:2020}%
      \BibitemOpen
      \bibfield  {author} {\bibinfo {author} {\bibfnamefont {R.}~\bibnamefont {Uppu}}, \bibinfo {author} {\bibfnamefont {F.~T.}\ \bibnamefont {Pedersen}}, \bibinfo {author} {\bibfnamefont {Y.}~\bibnamefont {Wang}}, \bibinfo {author} {\bibfnamefont {C.~T.}\ \bibnamefont {Olesen}}, \bibinfo {author} {\bibfnamefont {C.}~\bibnamefont {Papon}}, \bibinfo {author} {\bibfnamefont {X.}~\bibnamefont {Zhou}}, \bibinfo {author} {\bibfnamefont {L.}~\bibnamefont {Midolo}}, \bibinfo {author} {\bibfnamefont {S.}~\bibnamefont {Scholz}}, \bibinfo {author} {\bibfnamefont {A.~D.}\ \bibnamefont {Wieck}}, \bibinfo {author} {\bibfnamefont {A.}~\bibnamefont {Ludwig}},\ and\ \bibinfo {author} {\bibfnamefont {P.}~\bibnamefont {Lodahl}},\ }\bibfield  {title} {\bibinfo {title} {Scalable integrated single-photon source},\ }\href {https://doi.org/10.1126/sciadv.abc8268} {\bibfield  {journal} {\bibinfo  {journal} {Sci. Adv.}\ }\textbf {\bibinfo {volume} {6}},\ \bibinfo {pages} {eabc8268} (\bibinfo {year} {2020})}\BibitemShut {NoStop}%
    \bibitem [{\citenamefont {Kim}\ \emph {et~al.}(2020)\citenamefont {Kim}, \citenamefont {Aghaeimeibodi}, \citenamefont {Carolan}, \citenamefont {Englund},\ and\ \citenamefont {Waks}}]{Kim:2020}%
      \BibitemOpen
      \bibfield  {author} {\bibinfo {author} {\bibfnamefont {J.-H.}\ \bibnamefont {Kim}}, \bibinfo {author} {\bibfnamefont {S.}~\bibnamefont {Aghaeimeibodi}}, \bibinfo {author} {\bibfnamefont {J.}~\bibnamefont {Carolan}}, \bibinfo {author} {\bibfnamefont {D.}~\bibnamefont {Englund}},\ and\ \bibinfo {author} {\bibfnamefont {E.}~\bibnamefont {Waks}},\ }\bibfield  {title} {\bibinfo {title} {Hybrid integration methods for on-chip quantum photonics},\ }\href {https://doi.org/10.1364/optica.384118} {\bibfield  {journal} {\bibinfo  {journal} {Optica}\ }\textbf {\bibinfo {volume} {7}},\ \bibinfo {pages} {291} (\bibinfo {year} {2020})}\BibitemShut {NoStop}%
    \bibitem [{\citenamefont {Elshaari}\ \emph {et~al.}(2020)\citenamefont {Elshaari}, \citenamefont {Pernice}, \citenamefont {Srinivasan}, \citenamefont {Benson},\ and\ \citenamefont {Zwiller}}]{Elshaari:2020}%
      \BibitemOpen
      \bibfield  {author} {\bibinfo {author} {\bibfnamefont {A.~W.}\ \bibnamefont {Elshaari}}, \bibinfo {author} {\bibfnamefont {W.}~\bibnamefont {Pernice}}, \bibinfo {author} {\bibfnamefont {K.}~\bibnamefont {Srinivasan}}, \bibinfo {author} {\bibfnamefont {O.}~\bibnamefont {Benson}},\ and\ \bibinfo {author} {\bibfnamefont {V.}~\bibnamefont {Zwiller}},\ }\bibfield  {title} {\bibinfo {title} {Hybrid integrated quantum photonic circuits},\ }\href {https://doi.org/10.1038/s41566-020-0609-x} {\bibfield  {journal} {\bibinfo  {journal} {Nat. Photonics}\ }\textbf {\bibinfo {volume} {14}},\ \bibinfo {pages} {285} (\bibinfo {year} {2020})}\BibitemShut {NoStop}%
    \bibitem [{\citenamefont {Aharonovich}\ \emph {et~al.}(2022)\citenamefont {Aharonovich}, \citenamefont {Tetienne},\ and\ \citenamefont {Toth}}]{Aharonovich:2022}%
      \BibitemOpen
      \bibfield  {author} {\bibinfo {author} {\bibfnamefont {I.}~\bibnamefont {Aharonovich}}, \bibinfo {author} {\bibfnamefont {J.-P.}\ \bibnamefont {Tetienne}},\ and\ \bibinfo {author} {\bibfnamefont {M.}~\bibnamefont {Toth}},\ }\bibfield  {title} {\bibinfo {title} {Quantum emitters in hexagonal boron nitride},\ }\href {https://doi.org/10.1021/acs.nanolett.2c03743} {\bibfield  {journal} {\bibinfo  {journal} {Nano Lett.}\ }\textbf {\bibinfo {volume} {22}},\ \bibinfo {pages} {9227} (\bibinfo {year} {2022})}\BibitemShut {NoStop}%
    \bibitem [{\citenamefont {Turunen}\ \emph {et~al.}(2022)\citenamefont {Turunen}, \citenamefont {Brotons-Gisbert}, \citenamefont {Dai}, \citenamefont {Wang}, \citenamefont {Scerri}, \citenamefont {Bonato}, \citenamefont {Jöns}, \citenamefont {Sun},\ and\ \citenamefont {Gerardot}}]{Turunen:2022}%
      \BibitemOpen
      \bibfield  {author} {\bibinfo {author} {\bibfnamefont {M.}~\bibnamefont {Turunen}}, \bibinfo {author} {\bibfnamefont {M.}~\bibnamefont {Brotons-Gisbert}}, \bibinfo {author} {\bibfnamefont {Y.}~\bibnamefont {Dai}}, \bibinfo {author} {\bibfnamefont {Y.}~\bibnamefont {Wang}}, \bibinfo {author} {\bibfnamefont {E.}~\bibnamefont {Scerri}}, \bibinfo {author} {\bibfnamefont {C.}~\bibnamefont {Bonato}}, \bibinfo {author} {\bibfnamefont {K.~D.}\ \bibnamefont {Jöns}}, \bibinfo {author} {\bibfnamefont {Z.}~\bibnamefont {Sun}},\ and\ \bibinfo {author} {\bibfnamefont {B.~D.}\ \bibnamefont {Gerardot}},\ }\bibfield  {title} {\bibinfo {title} {Quantum photonics with layered 2d materials},\ }\href {https://doi.org/10.1038/s42254-021-00408-0} {\bibfield  {journal} {\bibinfo  {journal} {Nat. Rev. Phys.}\ }\textbf {\bibinfo {volume} {4}},\ \bibinfo {pages} {219} (\bibinfo {year} {2022})}\BibitemShut {NoStop}%
    \bibitem [{\citenamefont {Tran}\ \emph {et~al.}(2015)\citenamefont {Tran}, \citenamefont {Bray}, \citenamefont {Ford}, \citenamefont {Toth},\ and\ \citenamefont {Aharonovich}}]{Tran:2015}%
      \BibitemOpen
      \bibfield  {author} {\bibinfo {author} {\bibfnamefont {T.~T.}\ \bibnamefont {Tran}}, \bibinfo {author} {\bibfnamefont {K.}~\bibnamefont {Bray}}, \bibinfo {author} {\bibfnamefont {M.~J.}\ \bibnamefont {Ford}}, \bibinfo {author} {\bibfnamefont {M.}~\bibnamefont {Toth}},\ and\ \bibinfo {author} {\bibfnamefont {I.}~\bibnamefont {Aharonovich}},\ }\bibfield  {title} {\bibinfo {title} {Quantum emission from hexagonal boron nitride monolayers},\ }\href {https://doi.org/10.1038/nnano.2015.242} {\bibfield  {journal} {\bibinfo  {journal} {Nat. Nanotechnol.}\ }\textbf {\bibinfo {volume} {11}},\ \bibinfo {pages} {37} (\bibinfo {year} {2015})}\BibitemShut {NoStop}%
    \bibitem [{\citenamefont {Castelletto}\ \emph {et~al.}(2020)\citenamefont {Castelletto}, \citenamefont {Inam}, \citenamefont {Sato},\ and\ \citenamefont {Boretti}}]{Castelletto:2020}%
      \BibitemOpen
      \bibfield  {author} {\bibinfo {author} {\bibfnamefont {S.}~\bibnamefont {Castelletto}}, \bibinfo {author} {\bibfnamefont {F.~A.}\ \bibnamefont {Inam}}, \bibinfo {author} {\bibfnamefont {S.-i.}\ \bibnamefont {Sato}},\ and\ \bibinfo {author} {\bibfnamefont {A.}~\bibnamefont {Boretti}},\ }\bibfield  {title} {\bibinfo {title} {Hexagonal boron nitride: a review of the emerging material platform for single-photon sources and the spin–photon interface},\ }\href {https://doi.org/10.3762/bjnano.11.61} {\bibfield  {journal} {\bibinfo  {journal} {Beilstein J. Nanotechnol.}\ }\textbf {\bibinfo {volume} {11}},\ \bibinfo {pages} {740} (\bibinfo {year} {2020})}\BibitemShut {NoStop}%
    \bibitem [{\citenamefont {Shaik}\ and\ \citenamefont {Palla}(2021)}]{Shaik:2021}%
      \BibitemOpen
      \bibfield  {author} {\bibinfo {author} {\bibfnamefont {A.~B. D.-a.-j.-w.-i.}\ \bibnamefont {Shaik}}\ and\ \bibinfo {author} {\bibfnamefont {P.}~\bibnamefont {Palla}},\ }\bibfield  {title} {\bibinfo {title} {Optical quantum technologies with hexagonal boron nitride single photon sources},\ }\href {https://doi.org/10.1038/s41598-021-90804-4} {\bibfield  {journal} {\bibinfo  {journal} {Sci. Rep.}\ }\textbf {\bibinfo {volume} {11}},\ \bibinfo {pages} {12285} (\bibinfo {year} {2021})}\BibitemShut {NoStop}%
    \bibitem [{\citenamefont {Li}\ \emph {et~al.}(2019)\citenamefont {Li}, \citenamefont {Xu}, \citenamefont {Mendelson}, \citenamefont {Kianinia}, \citenamefont {Toth},\ and\ \citenamefont {Aharonovich}}]{Li:2019}%
      \BibitemOpen
      \bibfield  {author} {\bibinfo {author} {\bibfnamefont {C.}~\bibnamefont {Li}}, \bibinfo {author} {\bibfnamefont {Z.-Q.}\ \bibnamefont {Xu}}, \bibinfo {author} {\bibfnamefont {N.}~\bibnamefont {Mendelson}}, \bibinfo {author} {\bibfnamefont {M.}~\bibnamefont {Kianinia}}, \bibinfo {author} {\bibfnamefont {M.}~\bibnamefont {Toth}},\ and\ \bibinfo {author} {\bibfnamefont {I.}~\bibnamefont {Aharonovich}},\ }\bibfield  {title} {\bibinfo {title} {Purification of single-photon emission from hbn using post-processing treatments},\ }\href {https://doi.org/10.1515/nanoph-2019-0099} {\bibfield  {journal} {\bibinfo  {journal} {Nanophotonics}\ }\textbf {\bibinfo {volume} {8}},\ \bibinfo {pages} {2049} (\bibinfo {year} {2019})}\BibitemShut {NoStop}%
    \bibitem [{\citenamefont {Mendelson}\ \emph {et~al.}(2019)\citenamefont {Mendelson}, \citenamefont {Xu}, \citenamefont {Tran}, \citenamefont {Kianinia}, \citenamefont {Scott}, \citenamefont {Bradac}, \citenamefont {Aharonovich},\ and\ \citenamefont {Toth}}]{Mendelson:2019}%
      \BibitemOpen
      \bibfield  {author} {\bibinfo {author} {\bibfnamefont {N.}~\bibnamefont {Mendelson}}, \bibinfo {author} {\bibfnamefont {Z.-Q.}\ \bibnamefont {Xu}}, \bibinfo {author} {\bibfnamefont {T.~T.}\ \bibnamefont {Tran}}, \bibinfo {author} {\bibfnamefont {M.}~\bibnamefont {Kianinia}}, \bibinfo {author} {\bibfnamefont {J.}~\bibnamefont {Scott}}, \bibinfo {author} {\bibfnamefont {C.}~\bibnamefont {Bradac}}, \bibinfo {author} {\bibfnamefont {I.}~\bibnamefont {Aharonovich}},\ and\ \bibinfo {author} {\bibfnamefont {M.}~\bibnamefont {Toth}},\ }\bibfield  {title} {\bibinfo {title} {Engineering and tuning of quantum emitters in few-layer hexagonal boron nitride},\ }\href {https://doi.org/10.1021/acsnano.8b08511} {\bibfield  {journal} {\bibinfo  {journal} {ACS Nano}\ }\textbf {\bibinfo {volume} {13}},\ \bibinfo {pages} {3132} (\bibinfo {year} {2019})}\BibitemShut {NoStop}%
    \bibitem [{\citenamefont {Lyu}\ \emph {et~al.}(2020)\citenamefont {Lyu}, \citenamefont {Zhu}, \citenamefont {Gu}, \citenamefont {Qiao}, \citenamefont {Watanabe}, \citenamefont {Taniguchi},\ and\ \citenamefont {Ye}}]{Lyu:2020}%
      \BibitemOpen
      \bibfield  {author} {\bibinfo {author} {\bibfnamefont {C.}~\bibnamefont {Lyu}}, \bibinfo {author} {\bibfnamefont {Y.}~\bibnamefont {Zhu}}, \bibinfo {author} {\bibfnamefont {P.}~\bibnamefont {Gu}}, \bibinfo {author} {\bibfnamefont {J.}~\bibnamefont {Qiao}}, \bibinfo {author} {\bibfnamefont {K.}~\bibnamefont {Watanabe}}, \bibinfo {author} {\bibfnamefont {T.}~\bibnamefont {Taniguchi}},\ and\ \bibinfo {author} {\bibfnamefont {Y.}~\bibnamefont {Ye}},\ }\bibfield  {title} {\bibinfo {title} {Single-photon emission from two-dimensional hexagonal boron nitride annealed in a carbon-rich environment},\ }\href {https://doi.org/10.1063/5.0025792} {\bibfield  {journal} {\bibinfo  {journal} {Appl. Phys. Lett.}\ }\textbf {\bibinfo {volume} {117}},\ \bibinfo {pages} {244002} (\bibinfo {year} {2020})}\BibitemShut {NoStop}%
    \bibitem [{\citenamefont {Proscia}\ \emph {et~al.}(2018)\citenamefont {Proscia}, \citenamefont {Shotan}, \citenamefont {Jayakumar}, \citenamefont {Reddy}, \citenamefont {Cohen}, \citenamefont {Dollar}, \citenamefont {Alkauskas}, \citenamefont {Doherty}, \citenamefont {Meriles},\ and\ \citenamefont {Menon}}]{Proscia:2018}%
      \BibitemOpen
      \bibfield  {author} {\bibinfo {author} {\bibfnamefont {N.~V.}\ \bibnamefont {Proscia}}, \bibinfo {author} {\bibfnamefont {Z.}~\bibnamefont {Shotan}}, \bibinfo {author} {\bibfnamefont {H.}~\bibnamefont {Jayakumar}}, \bibinfo {author} {\bibfnamefont {P.}~\bibnamefont {Reddy}}, \bibinfo {author} {\bibfnamefont {C.}~\bibnamefont {Cohen}}, \bibinfo {author} {\bibfnamefont {M.}~\bibnamefont {Dollar}}, \bibinfo {author} {\bibfnamefont {A.}~\bibnamefont {Alkauskas}}, \bibinfo {author} {\bibfnamefont {M.}~\bibnamefont {Doherty}}, \bibinfo {author} {\bibfnamefont {C.~A.}\ \bibnamefont {Meriles}},\ and\ \bibinfo {author} {\bibfnamefont {V.~M.}\ \bibnamefont {Menon}},\ }\bibfield  {title} {\bibinfo {title} {Near-deterministic activation of room-temperature quantum emitters in hexagonal boron nitride},\ }\href {https://doi.org/10.1364/optica.5.001128} {\bibfield  {journal} {\bibinfo  {journal} {Optica}\ }\textbf {\bibinfo {volume} {5}},\ \bibinfo {pages} {1128} (\bibinfo {year} {2018})}\BibitemShut {NoStop}%
    \bibitem [{\citenamefont {Li}\ \emph {et~al.}(2021)\citenamefont {Li}, \citenamefont {Mendelson}, \citenamefont {Ritika}, \citenamefont {Chen}, \citenamefont {Xu}, \citenamefont {Toth},\ and\ \citenamefont {Aharonovich}}]{Li:2021}%
      \BibitemOpen
      \bibfield  {author} {\bibinfo {author} {\bibfnamefont {C.}~\bibnamefont {Li}}, \bibinfo {author} {\bibfnamefont {N.}~\bibnamefont {Mendelson}}, \bibinfo {author} {\bibfnamefont {R.}~\bibnamefont {Ritika}}, \bibinfo {author} {\bibfnamefont {Y.}~\bibnamefont {Chen}}, \bibinfo {author} {\bibfnamefont {Z.-Q.}\ \bibnamefont {Xu}}, \bibinfo {author} {\bibfnamefont {M.}~\bibnamefont {Toth}},\ and\ \bibinfo {author} {\bibfnamefont {I.}~\bibnamefont {Aharonovich}},\ }\bibfield  {title} {\bibinfo {title} {Scalable and deterministic fabrication of quantum emitter arrays from hexagonal boron nitride},\ }\href {https://doi.org/10.1021/acs.nanolett.1c00685} {\bibfield  {journal} {\bibinfo  {journal} {Nano Lett.}\ }\textbf {\bibinfo {volume} {21}},\ \bibinfo {pages} {3626} (\bibinfo {year} {2021})}\BibitemShut {NoStop}%
    \bibitem [{\citenamefont {Xu}\ \emph {et~al.}(2021)\citenamefont {Xu}, \citenamefont {Martin}, \citenamefont {Sychev}, \citenamefont {Lagutchev}, \citenamefont {Chen}, \citenamefont {Taniguchi}, \citenamefont {Watanabe}, \citenamefont {Shalaev},\ and\ \citenamefont {Boltasseva}}]{Xu:2021}%
      \BibitemOpen
      \bibfield  {author} {\bibinfo {author} {\bibfnamefont {X.}~\bibnamefont {Xu}}, \bibinfo {author} {\bibfnamefont {Z.~O.}\ \bibnamefont {Martin}}, \bibinfo {author} {\bibfnamefont {D.}~\bibnamefont {Sychev}}, \bibinfo {author} {\bibfnamefont {A.~S.}\ \bibnamefont {Lagutchev}}, \bibinfo {author} {\bibfnamefont {Y.~P.}\ \bibnamefont {Chen}}, \bibinfo {author} {\bibfnamefont {T.}~\bibnamefont {Taniguchi}}, \bibinfo {author} {\bibfnamefont {K.}~\bibnamefont {Watanabe}}, \bibinfo {author} {\bibfnamefont {V.~M.}\ \bibnamefont {Shalaev}},\ and\ \bibinfo {author} {\bibfnamefont {A.}~\bibnamefont {Boltasseva}},\ }\bibfield  {title} {\bibinfo {title} {Creating quantum emitters in hexagonal boron nitride deterministically on chip-compatible substrates},\ }\href {https://doi.org/10.1021/acs.nanolett.1c02640} {\bibfield  {journal} {\bibinfo  {journal} {Nano Lett.}\ }\textbf {\bibinfo {volume} {21}},\ \bibinfo {pages} {8182} (\bibinfo {year} {2021})}\BibitemShut {NoStop}%
    \bibitem [{\citenamefont {Choi}\ \emph {et~al.}(2016)\citenamefont {Choi}, \citenamefont {Tran}, \citenamefont {Elbadawi}, \citenamefont {Lobo}, \citenamefont {Wang}, \citenamefont {Juodkazis}, \citenamefont {Seniutinas}, \citenamefont {Toth},\ and\ \citenamefont {Aharonovich}}]{Choi:2016}%
      \BibitemOpen
      \bibfield  {author} {\bibinfo {author} {\bibfnamefont {S.}~\bibnamefont {Choi}}, \bibinfo {author} {\bibfnamefont {T.~T.}\ \bibnamefont {Tran}}, \bibinfo {author} {\bibfnamefont {C.}~\bibnamefont {Elbadawi}}, \bibinfo {author} {\bibfnamefont {C.}~\bibnamefont {Lobo}}, \bibinfo {author} {\bibfnamefont {X.}~\bibnamefont {Wang}}, \bibinfo {author} {\bibfnamefont {S.}~\bibnamefont {Juodkazis}}, \bibinfo {author} {\bibfnamefont {G.}~\bibnamefont {Seniutinas}}, \bibinfo {author} {\bibfnamefont {M.}~\bibnamefont {Toth}},\ and\ \bibinfo {author} {\bibfnamefont {I.}~\bibnamefont {Aharonovich}},\ }\bibfield  {title} {\bibinfo {title} {Engineering and localization of quantum emitters in large hexagonal boron nitride layers},\ }\href {https://doi.org/10.1021/acsami.6b09875} {\bibfield  {journal} {\bibinfo  {journal} {ACS Appl. Mater. Interfaces.}\ }\textbf {\bibinfo {volume} {8}},\ \bibinfo {pages} {29642} (\bibinfo {year} {2016})}\BibitemShut {NoStop}%
    \bibitem [{\citenamefont {Xu}\ \emph {et~al.}(2018)\citenamefont {Xu}, \citenamefont {Elbadawi}, \citenamefont {Tran}, \citenamefont {Kianinia}, \citenamefont {Li}, \citenamefont {Liu}, \citenamefont {Hoffman}, \citenamefont {Nguyen}, \citenamefont {Kim}, \citenamefont {Edgar}, \citenamefont {Wu}, \citenamefont {Song}, \citenamefont {Ali}, \citenamefont {Ford}, \citenamefont {Toth},\ and\ \citenamefont {Aharonovich}}]{Xu:2018}%
      \BibitemOpen
      \bibfield  {author} {\bibinfo {author} {\bibfnamefont {Z.-Q.}\ \bibnamefont {Xu}}, \bibinfo {author} {\bibfnamefont {C.}~\bibnamefont {Elbadawi}}, \bibinfo {author} {\bibfnamefont {T.~T.}\ \bibnamefont {Tran}}, \bibinfo {author} {\bibfnamefont {M.}~\bibnamefont {Kianinia}}, \bibinfo {author} {\bibfnamefont {X.}~\bibnamefont {Li}}, \bibinfo {author} {\bibfnamefont {D.}~\bibnamefont {Liu}}, \bibinfo {author} {\bibfnamefont {T.~B.}\ \bibnamefont {Hoffman}}, \bibinfo {author} {\bibfnamefont {M.}~\bibnamefont {Nguyen}}, \bibinfo {author} {\bibfnamefont {S.}~\bibnamefont {Kim}}, \bibinfo {author} {\bibfnamefont {J.~H.}\ \bibnamefont {Edgar}}, \bibinfo {author} {\bibfnamefont {X.}~\bibnamefont {Wu}}, \bibinfo {author} {\bibfnamefont {L.}~\bibnamefont {Song}}, \bibinfo {author} {\bibfnamefont {S.}~\bibnamefont {Ali}}, \bibinfo {author} {\bibfnamefont {M.}~\bibnamefont {Ford}}, \bibinfo {author} {\bibfnamefont {M.}~\bibnamefont {Toth}},\ and\ \bibinfo {author} {\bibfnamefont {I.}~\bibnamefont {Aharonovich}},\ }\bibfield  {title} {\bibinfo {title} {Single photon emission from plasma treated 2d hexagonal boron nitride},\ }\href {https://doi.org/10.1039/c7nr08222c} {\bibfield  {journal} {\bibinfo  {journal} {Nanoscale}\ }\textbf {\bibinfo {volume} {10}},\ \bibinfo {pages} {7957} (\bibinfo {year} {2018})}\BibitemShut {NoStop}%
    \bibitem [{\citenamefont {Vogl}\ \emph {et~al.}(2018)\citenamefont {Vogl}, \citenamefont {Campbell}, \citenamefont {Buchler}, \citenamefont {Lu},\ and\ \citenamefont {Lam}}]{Vogl:2018}%
      \BibitemOpen
      \bibfield  {author} {\bibinfo {author} {\bibfnamefont {T.}~\bibnamefont {Vogl}}, \bibinfo {author} {\bibfnamefont {G.}~\bibnamefont {Campbell}}, \bibinfo {author} {\bibfnamefont {B.~C.}\ \bibnamefont {Buchler}}, \bibinfo {author} {\bibfnamefont {Y.}~\bibnamefont {Lu}},\ and\ \bibinfo {author} {\bibfnamefont {P.~K.}\ \bibnamefont {Lam}},\ }\bibfield  {title} {\bibinfo {title} {Fabrication and deterministic transfer of high-quality quantum emitters in hexagonal boron nitride},\ }\href {https://doi.org/10.1021/acsphotonics.8b00127} {\bibfield  {journal} {\bibinfo  {journal} {ACS Photonics}\ }\textbf {\bibinfo {volume} {5}},\ \bibinfo {pages} {2305} (\bibinfo {year} {2018})}\BibitemShut {NoStop}%
    \bibitem [{\citenamefont {Gu}\ \emph {et~al.}(2021)\citenamefont {Gu}, \citenamefont {Wang}, \citenamefont {Zhu}, \citenamefont {Han}, \citenamefont {Bai}, \citenamefont {Zhang}, \citenamefont {Li}, \citenamefont {Qin}, \citenamefont {Liu}, \citenamefont {Guo}, \citenamefont {Shan}, \citenamefont {Xiong}, \citenamefont {Gao}, \citenamefont {He}, \citenamefont {Han}, \citenamefont {Liu},\ and\ \citenamefont {Zhao}}]{Gu:2021}%
      \BibitemOpen
      \bibfield  {author} {\bibinfo {author} {\bibfnamefont {R.}~\bibnamefont {Gu}}, \bibinfo {author} {\bibfnamefont {L.}~\bibnamefont {Wang}}, \bibinfo {author} {\bibfnamefont {H.}~\bibnamefont {Zhu}}, \bibinfo {author} {\bibfnamefont {S.}~\bibnamefont {Han}}, \bibinfo {author} {\bibfnamefont {Y.}~\bibnamefont {Bai}}, \bibinfo {author} {\bibfnamefont {X.}~\bibnamefont {Zhang}}, \bibinfo {author} {\bibfnamefont {B.}~\bibnamefont {Li}}, \bibinfo {author} {\bibfnamefont {C.}~\bibnamefont {Qin}}, \bibinfo {author} {\bibfnamefont {J.}~\bibnamefont {Liu}}, \bibinfo {author} {\bibfnamefont {G.}~\bibnamefont {Guo}}, \bibinfo {author} {\bibfnamefont {X.}~\bibnamefont {Shan}}, \bibinfo {author} {\bibfnamefont {G.}~\bibnamefont {Xiong}}, \bibinfo {author} {\bibfnamefont {J.}~\bibnamefont {Gao}}, \bibinfo {author} {\bibfnamefont {C.}~\bibnamefont {He}}, \bibinfo {author} {\bibfnamefont {Z.}~\bibnamefont {Han}}, \bibinfo {author} {\bibfnamefont {X.}~\bibnamefont {Liu}},\ and\ \bibinfo {author} {\bibfnamefont {F.}~\bibnamefont {Zhao}},\ }\bibfield  {title} {\bibinfo {title} {Engineering and microscopic mechanism of quantum emitters induced by heavy ions in hbn},\ }\href {https://doi.org/10.1021/acsphotonics.1c00364} {\bibfield  {journal} {\bibinfo  {journal} {ACS Photonics}\ }\textbf {\bibinfo {volume} {8}},\ \bibinfo {pages} {2912} (\bibinfo {year} {2021})}\BibitemShut {NoStop}%
    \bibitem [{\citenamefont {Fischer}\ \emph {et~al.}(2021)\citenamefont {Fischer}, \citenamefont {Caridad}, \citenamefont {Sajid}, \citenamefont {Ghaderzadeh}, \citenamefont {Ghorbani-Asl}, \citenamefont {Gammelgaard}, \citenamefont {Bøggild}, \citenamefont {Thygesen}, \citenamefont {Krasheninnikov}, \citenamefont {Xiao}, \citenamefont {Wubs},\ and\ \citenamefont {Stenger}}]{Fischer:2021}%
      \BibitemOpen
      \bibfield  {author} {\bibinfo {author} {\bibfnamefont {M.}~\bibnamefont {Fischer}}, \bibinfo {author} {\bibfnamefont {J.~M.}\ \bibnamefont {Caridad}}, \bibinfo {author} {\bibfnamefont {A.}~\bibnamefont {Sajid}}, \bibinfo {author} {\bibfnamefont {S.}~\bibnamefont {Ghaderzadeh}}, \bibinfo {author} {\bibfnamefont {M.}~\bibnamefont {Ghorbani-Asl}}, \bibinfo {author} {\bibfnamefont {L.}~\bibnamefont {Gammelgaard}}, \bibinfo {author} {\bibfnamefont {P.}~\bibnamefont {Bøggild}}, \bibinfo {author} {\bibfnamefont {K.~S.}\ \bibnamefont {Thygesen}}, \bibinfo {author} {\bibfnamefont {A.~V.}\ \bibnamefont {Krasheninnikov}}, \bibinfo {author} {\bibfnamefont {S.}~\bibnamefont {Xiao}}, \bibinfo {author} {\bibfnamefont {M.}~\bibnamefont {Wubs}},\ and\ \bibinfo {author} {\bibfnamefont {N.}~\bibnamefont {Stenger}},\ }\bibfield  {title} {\bibinfo {title} {Controlled generation of luminescent centers in hexagonal boron nitride by irradiation engineering},\ }\href {https://doi.org/10.1126/sciadv.abe7138} {\bibfield  {journal} {\bibinfo  {journal} {Sci. Adv.}\ }\textbf {\bibinfo {volume} {7}},\ \bibinfo {pages} {eabe7138} (\bibinfo {year} {2021})}\BibitemShut {NoStop}%
    \bibitem [{\citenamefont {Ziegler}\ \emph {et~al.}(2019)\citenamefont {Ziegler}, \citenamefont {Klaiss}, \citenamefont {Blaikie}, \citenamefont {Miller}, \citenamefont {Horowitz},\ and\ \citenamefont {Alemán}}]{Ziegler:2019}%
      \BibitemOpen
      \bibfield  {author} {\bibinfo {author} {\bibfnamefont {J.}~\bibnamefont {Ziegler}}, \bibinfo {author} {\bibfnamefont {R.}~\bibnamefont {Klaiss}}, \bibinfo {author} {\bibfnamefont {A.}~\bibnamefont {Blaikie}}, \bibinfo {author} {\bibfnamefont {D.}~\bibnamefont {Miller}}, \bibinfo {author} {\bibfnamefont {V.~R.}\ \bibnamefont {Horowitz}},\ and\ \bibinfo {author} {\bibfnamefont {B.~J.}\ \bibnamefont {Alemán}},\ }\bibfield  {title} {\bibinfo {title} {Deterministic quantum emitter formation in hexagonal boron nitride via controlled edge creation},\ }\href {https://doi.org/10.1021/acs.nanolett.9b00357} {\bibfield  {journal} {\bibinfo  {journal} {Nano Lett.}\ }\textbf {\bibinfo {volume} {19}},\ \bibinfo {pages} {2121} (\bibinfo {year} {2019})}\BibitemShut {NoStop}%
    \bibitem [{\citenamefont {Glushkov}\ \emph {et~al.}(2022)\citenamefont {Glushkov}, \citenamefont {Macha}, \citenamefont {Räth}, \citenamefont {Navikas}, \citenamefont {Ronceray}, \citenamefont {Cheon}, \citenamefont {Ahmed}, \citenamefont {Avsar}, \citenamefont {Watanabe}, \citenamefont {Taniguchi}, \citenamefont {Shorubalko}, \citenamefont {Kis}, \citenamefont {Fantner},\ and\ \citenamefont {Radenovic}}]{Glushkov:2022}%
      \BibitemOpen
      \bibfield  {author} {\bibinfo {author} {\bibfnamefont {E.}~\bibnamefont {Glushkov}}, \bibinfo {author} {\bibfnamefont {M.}~\bibnamefont {Macha}}, \bibinfo {author} {\bibfnamefont {E.}~\bibnamefont {Räth}}, \bibinfo {author} {\bibfnamefont {V.}~\bibnamefont {Navikas}}, \bibinfo {author} {\bibfnamefont {N.}~\bibnamefont {Ronceray}}, \bibinfo {author} {\bibfnamefont {C.~Y.}\ \bibnamefont {Cheon}}, \bibinfo {author} {\bibfnamefont {A.}~\bibnamefont {Ahmed}}, \bibinfo {author} {\bibfnamefont {A.}~\bibnamefont {Avsar}}, \bibinfo {author} {\bibfnamefont {K.}~\bibnamefont {Watanabe}}, \bibinfo {author} {\bibfnamefont {T.}~\bibnamefont {Taniguchi}}, \bibinfo {author} {\bibfnamefont {I.}~\bibnamefont {Shorubalko}}, \bibinfo {author} {\bibfnamefont {A.}~\bibnamefont {Kis}}, \bibinfo {author} {\bibfnamefont {G.}~\bibnamefont {Fantner}},\ and\ \bibinfo {author} {\bibfnamefont {A.}~\bibnamefont {Radenovic}},\ }\bibfield  {title} {\bibinfo {title} {Engineering optically active defects in hexagonal boron nitride using focused ion beam and water},\ }\href {https://doi.org/10.1021/acsnano.1c07086} {\bibfield  {journal} {\bibinfo  {journal} {ACS Nano}\ }\textbf {\bibinfo {volume} {16}},\ \bibinfo {pages} {3695} (\bibinfo {year} {2022})}\BibitemShut {NoStop}%
    \bibitem [{\citenamefont {Ngoc My~Duong}\ \emph {et~al.}(2018)\citenamefont {Ngoc My~Duong}, \citenamefont {Nguyen}, \citenamefont {Kianinia}, \citenamefont {Ohshima}, \citenamefont {Abe}, \citenamefont {Watanabe}, \citenamefont {Taniguchi}, \citenamefont {Edgar}, \citenamefont {Aharonovich},\ and\ \citenamefont {Toth}}]{NgocMyDuong:2018}%
      \BibitemOpen
      \bibfield  {author} {\bibinfo {author} {\bibfnamefont {H.}~\bibnamefont {Ngoc My~Duong}}, \bibinfo {author} {\bibfnamefont {M.~A.~P.}\ \bibnamefont {Nguyen}}, \bibinfo {author} {\bibfnamefont {M.}~\bibnamefont {Kianinia}}, \bibinfo {author} {\bibfnamefont {T.}~\bibnamefont {Ohshima}}, \bibinfo {author} {\bibfnamefont {H.}~\bibnamefont {Abe}}, \bibinfo {author} {\bibfnamefont {K.}~\bibnamefont {Watanabe}}, \bibinfo {author} {\bibfnamefont {T.}~\bibnamefont {Taniguchi}}, \bibinfo {author} {\bibfnamefont {J.~H.}\ \bibnamefont {Edgar}}, \bibinfo {author} {\bibfnamefont {I.}~\bibnamefont {Aharonovich}},\ and\ \bibinfo {author} {\bibfnamefont {M.}~\bibnamefont {Toth}},\ }\bibfield  {title} {\bibinfo {title} {Effects of high-energy electron irradiation on quantum emitters in hexagonal boron nitride},\ }\href {https://doi.org/10.1021/acsami.8b07506} {\bibfield  {journal} {\bibinfo  {journal} {ACS Appl. Mater. Interfaces.}\ }\textbf {\bibinfo {volume} {10}},\ \bibinfo {pages} {24886} (\bibinfo {year} {2018})}\BibitemShut {NoStop}%
    \bibitem [{\citenamefont {Fournier}\ \emph {et~al.}(2021)\citenamefont {Fournier}, \citenamefont {Plaud}, \citenamefont {Roux}, \citenamefont {Pierret}, \citenamefont {Rosticher}, \citenamefont {Watanabe}, \citenamefont {Taniguchi}, \citenamefont {Buil}, \citenamefont {Quélin}, \citenamefont {Barjon}, \citenamefont {Hermier},\ and\ \citenamefont {Delteil}}]{Fournier:2021}%
      \BibitemOpen
      \bibfield  {author} {\bibinfo {author} {\bibfnamefont {C.}~\bibnamefont {Fournier}}, \bibinfo {author} {\bibfnamefont {A.}~\bibnamefont {Plaud}}, \bibinfo {author} {\bibfnamefont {S.}~\bibnamefont {Roux}}, \bibinfo {author} {\bibfnamefont {A.}~\bibnamefont {Pierret}}, \bibinfo {author} {\bibfnamefont {M.}~\bibnamefont {Rosticher}}, \bibinfo {author} {\bibfnamefont {K.}~\bibnamefont {Watanabe}}, \bibinfo {author} {\bibfnamefont {T.}~\bibnamefont {Taniguchi}}, \bibinfo {author} {\bibfnamefont {S.}~\bibnamefont {Buil}}, \bibinfo {author} {\bibfnamefont {X.}~\bibnamefont {Quélin}}, \bibinfo {author} {\bibfnamefont {J.}~\bibnamefont {Barjon}}, \bibinfo {author} {\bibfnamefont {J.-P.}\ \bibnamefont {Hermier}},\ and\ \bibinfo {author} {\bibfnamefont {A.}~\bibnamefont {Delteil}},\ }\bibfield  {title} {\bibinfo {title} {Position-controlled quantum emitters with reproducible emission wavelength in hexagonal boron nitride},\ }\href {https://doi.org/10.1038/s41467-021-24019-6} {\bibfield  {journal} {\bibinfo  {journal} {Nat. Commun.}\ }\textbf {\bibinfo {volume} {12}},\ \bibinfo {pages} {3779} (\bibinfo {year} {2021})}\BibitemShut {NoStop}%
    \bibitem [{\citenamefont {Hou}\ \emph {et~al.}(2017)\citenamefont {Hou}, \citenamefont {Birowosuto}, \citenamefont {Umar}, \citenamefont {Anicet}, \citenamefont {Tay}, \citenamefont {Coquet}, \citenamefont {Tay}, \citenamefont {Wang},\ and\ \citenamefont {Teo}}]{Hou:2017}%
      \BibitemOpen
      \bibfield  {author} {\bibinfo {author} {\bibfnamefont {S.}~\bibnamefont {Hou}}, \bibinfo {author} {\bibfnamefont {M.~D.}\ \bibnamefont {Birowosuto}}, \bibinfo {author} {\bibfnamefont {S.}~\bibnamefont {Umar}}, \bibinfo {author} {\bibfnamefont {M.~A.}\ \bibnamefont {Anicet}}, \bibinfo {author} {\bibfnamefont {R.~Y.}\ \bibnamefont {Tay}}, \bibinfo {author} {\bibfnamefont {P.}~\bibnamefont {Coquet}}, \bibinfo {author} {\bibfnamefont {B.~K.}\ \bibnamefont {Tay}}, \bibinfo {author} {\bibfnamefont {H.}~\bibnamefont {Wang}},\ and\ \bibinfo {author} {\bibfnamefont {E.~H.~T.}\ \bibnamefont {Teo}},\ }\bibfield  {title} {\bibinfo {title} {Localized emission from laser-irradiated defects in 2d hexagonal boron nitride},\ }\href {https://doi.org/10.1088/2053-1583/aa8e61} {\bibfield  {journal} {\bibinfo  {journal} {2D Materials}\ }\textbf {\bibinfo {volume} {5}},\ \bibinfo {pages} {015010} (\bibinfo {year} {2017})}\BibitemShut {NoStop}%
    \bibitem [{\citenamefont {Gan}\ \emph {et~al.}(2022)\citenamefont {Gan}, \citenamefont {Zhang}, \citenamefont {Zhang}, \citenamefont {Zhang}, \citenamefont {Sun}, \citenamefont {Li},\ and\ \citenamefont {Ning}}]{Gan:2022}%
      \BibitemOpen
      \bibfield  {author} {\bibinfo {author} {\bibfnamefont {L.}~\bibnamefont {Gan}}, \bibinfo {author} {\bibfnamefont {D.}~\bibnamefont {Zhang}}, \bibinfo {author} {\bibfnamefont {R.}~\bibnamefont {Zhang}}, \bibinfo {author} {\bibfnamefont {Q.}~\bibnamefont {Zhang}}, \bibinfo {author} {\bibfnamefont {H.}~\bibnamefont {Sun}}, \bibinfo {author} {\bibfnamefont {Y.}~\bibnamefont {Li}},\ and\ \bibinfo {author} {\bibfnamefont {C.-Z.}\ \bibnamefont {Ning}},\ }\bibfield  {title} {\bibinfo {title} {Large-scale, high-yield laser fabrication of bright and pure single-photon emitters at room temperature in hexagonal boron nitride},\ }\href {https://doi.org/10.1021/acsnano.2c04386} {\bibfield  {journal} {\bibinfo  {journal} {ACS Nano}\ }\textbf {\bibinfo {volume} {16}},\ \bibinfo {pages} {14254} (\bibinfo {year} {2022})}\BibitemShut {NoStop}%
    \bibitem [{\citenamefont {Wang}\ \emph {et~al.}(2024)\citenamefont {Wang}, \citenamefont {Fang}, \citenamefont {Li}, \citenamefont {Wang},\ and\ \citenamefont {Sun}}]{Wang:2024}%
      \BibitemOpen
      \bibfield  {author} {\bibinfo {author} {\bibfnamefont {X.-J.}\ \bibnamefont {Wang}}, \bibinfo {author} {\bibfnamefont {H.-H.}\ \bibnamefont {Fang}}, \bibinfo {author} {\bibfnamefont {Z.-Z.}\ \bibnamefont {Li}}, \bibinfo {author} {\bibfnamefont {D.}~\bibnamefont {Wang}},\ and\ \bibinfo {author} {\bibfnamefont {H.-B.}\ \bibnamefont {Sun}},\ }\bibfield  {title} {\bibinfo {title} {Laser manufacturing of spatial resolution approaching quantum limit},\ }\href {https://doi.org/10.1038/s41377-023-01354-5} {\bibfield  {journal} {\bibinfo  {journal} {Light: Sci. Appl.}\ }\textbf {\bibinfo {volume} {13}},\ \bibinfo {pages} {6} (\bibinfo {year} {2024})}\BibitemShut {NoStop}%
    \bibitem [{\citenamefont {Su}\ \emph {et~al.}(2024)\citenamefont {Su}, \citenamefont {Janzen}, \citenamefont {He}, \citenamefont {Li}, \citenamefont {Zettl}, \citenamefont {Caldwell}, \citenamefont {Edgar},\ and\ \citenamefont {Aharonovich}}]{Su:2024}%
      \BibitemOpen
      \bibfield  {author} {\bibinfo {author} {\bibfnamefont {C.}~\bibnamefont {Su}}, \bibinfo {author} {\bibfnamefont {E.}~\bibnamefont {Janzen}}, \bibinfo {author} {\bibfnamefont {M.}~\bibnamefont {He}}, \bibinfo {author} {\bibfnamefont {C.}~\bibnamefont {Li}}, \bibinfo {author} {\bibfnamefont {A.}~\bibnamefont {Zettl}}, \bibinfo {author} {\bibfnamefont {J.~D.}\ \bibnamefont {Caldwell}}, \bibinfo {author} {\bibfnamefont {J.~H.}\ \bibnamefont {Edgar}},\ and\ \bibinfo {author} {\bibfnamefont {I.}~\bibnamefont {Aharonovich}},\ }\bibfield  {title} {\bibinfo {title} {Fundamentals and emerging optical applications of hexagonal boron nitride: a tutorial},\ }\href {https://doi.org/10.1364/aop.502922} {\bibfield  {journal} {\bibinfo  {journal} {Adv. Opt. Photonics}\ }\textbf {\bibinfo {volume} {16}},\ \bibinfo {pages} {229} (\bibinfo {year} {2024})}\BibitemShut {NoStop}%
    \bibitem [{\citenamefont {Kim}\ \emph {et~al.}(2019)\citenamefont {Kim}, \citenamefont {Duong}, \citenamefont {Nguyen}, \citenamefont {Lu}, \citenamefont {Kianinia}, \citenamefont {Mendelson}, \citenamefont {Solntsev}, \citenamefont {Bradac}, \citenamefont {Englund},\ and\ \citenamefont {Aharonovich}}]{Kim:2019}%
      \BibitemOpen
      \bibfield  {author} {\bibinfo {author} {\bibfnamefont {S.}~\bibnamefont {Kim}}, \bibinfo {author} {\bibfnamefont {N.~M.~H.}\ \bibnamefont {Duong}}, \bibinfo {author} {\bibfnamefont {M.}~\bibnamefont {Nguyen}}, \bibinfo {author} {\bibfnamefont {T.}~\bibnamefont {Lu}}, \bibinfo {author} {\bibfnamefont {M.}~\bibnamefont {Kianinia}}, \bibinfo {author} {\bibfnamefont {N.}~\bibnamefont {Mendelson}}, \bibinfo {author} {\bibfnamefont {A.}~\bibnamefont {Solntsev}}, \bibinfo {author} {\bibfnamefont {C.}~\bibnamefont {Bradac}}, \bibinfo {author} {\bibfnamefont {D.~R.}\ \bibnamefont {Englund}},\ and\ \bibinfo {author} {\bibfnamefont {I.}~\bibnamefont {Aharonovich}},\ }\bibfield  {title} {\bibinfo {title} {Integrated on chip platform with quantum emitters in layered materials},\ }\href {https://doi.org/10.1002/adom.201901132} {\bibfield  {journal} {\bibinfo  {journal} {Adv. Opt. Mater.}\ }\textbf {\bibinfo {volume} {7}},\ \bibinfo {pages} {1901132} (\bibinfo {year} {2019})}\BibitemShut {NoStop}%
    \bibitem [{\citenamefont {Proscia}\ \emph {et~al.}(2020)\citenamefont {Proscia}, \citenamefont {Jayakumar}, \citenamefont {Ge}, \citenamefont {Lopez-Morales}, \citenamefont {Shotan}, \citenamefont {Zhou}, \citenamefont {Meriles},\ and\ \citenamefont {Menon}}]{Proscia:2020}%
      \BibitemOpen
      \bibfield  {author} {\bibinfo {author} {\bibfnamefont {N.~V.}\ \bibnamefont {Proscia}}, \bibinfo {author} {\bibfnamefont {H.}~\bibnamefont {Jayakumar}}, \bibinfo {author} {\bibfnamefont {X.}~\bibnamefont {Ge}}, \bibinfo {author} {\bibfnamefont {G.}~\bibnamefont {Lopez-Morales}}, \bibinfo {author} {\bibfnamefont {Z.}~\bibnamefont {Shotan}}, \bibinfo {author} {\bibfnamefont {W.}~\bibnamefont {Zhou}}, \bibinfo {author} {\bibfnamefont {C.~A.}\ \bibnamefont {Meriles}},\ and\ \bibinfo {author} {\bibfnamefont {V.~M.}\ \bibnamefont {Menon}},\ }\bibfield  {title} {\bibinfo {title} {Microcavity-coupled emitters in hexagonal boron nitride},\ }\href {https://doi.org/10.1515/nanoph-2020-0187} {\bibfield  {journal} {\bibinfo  {journal} {Nanophotonics}\ }\textbf {\bibinfo {volume} {9}},\ \bibinfo {pages} {2937} (\bibinfo {year} {2020})}\BibitemShut {NoStop}%
    \bibitem [{\citenamefont {Fröch}\ \emph {et~al.}(2020)\citenamefont {Fröch}, \citenamefont {Kim}, \citenamefont {Mendelson}, \citenamefont {Kianinia}, \citenamefont {Toth},\ and\ \citenamefont {Aharonovich}}]{Froech:2020}%
      \BibitemOpen
      \bibfield  {author} {\bibinfo {author} {\bibfnamefont {J.~E.}\ \bibnamefont {Fröch}}, \bibinfo {author} {\bibfnamefont {S.}~\bibnamefont {Kim}}, \bibinfo {author} {\bibfnamefont {N.}~\bibnamefont {Mendelson}}, \bibinfo {author} {\bibfnamefont {M.}~\bibnamefont {Kianinia}}, \bibinfo {author} {\bibfnamefont {M.}~\bibnamefont {Toth}},\ and\ \bibinfo {author} {\bibfnamefont {I.}~\bibnamefont {Aharonovich}},\ }\bibfield  {title} {\bibinfo {title} {Coupling hexagonal boron nitride quantum emitters to photonic crystal cavities},\ }\href {https://doi.org/10.1021/acsnano.0c01818} {\bibfield  {journal} {\bibinfo  {journal} {ACS Nano}\ }\textbf {\bibinfo {volume} {14}},\ \bibinfo {pages} {7085} (\bibinfo {year} {2020})}\BibitemShut {NoStop}%
    \bibitem [{\citenamefont {Parto}\ \emph {et~al.}(2022)\citenamefont {Parto}, \citenamefont {Azzam}, \citenamefont {Lewis}, \citenamefont {Patel}, \citenamefont {Umezawa}, \citenamefont {Watanabe}, \citenamefont {Taniguchi},\ and\ \citenamefont {Moody}}]{Parto:2022}%
      \BibitemOpen
      \bibfield  {author} {\bibinfo {author} {\bibfnamefont {K.}~\bibnamefont {Parto}}, \bibinfo {author} {\bibfnamefont {S.~I.}\ \bibnamefont {Azzam}}, \bibinfo {author} {\bibfnamefont {N.}~\bibnamefont {Lewis}}, \bibinfo {author} {\bibfnamefont {S.~D.}\ \bibnamefont {Patel}}, \bibinfo {author} {\bibfnamefont {S.}~\bibnamefont {Umezawa}}, \bibinfo {author} {\bibfnamefont {K.}~\bibnamefont {Watanabe}}, \bibinfo {author} {\bibfnamefont {T.}~\bibnamefont {Taniguchi}},\ and\ \bibinfo {author} {\bibfnamefont {G.}~\bibnamefont {Moody}},\ }\bibfield  {title} {\bibinfo {title} {Cavity-enhanced 2d material quantum emitters deterministically integrated with silicon nitride microresonators},\ }\href {https://doi.org/10.1021/acs.nanolett.2c03151} {\bibfield  {journal} {\bibinfo  {journal} {Nano Lett.}\ }\textbf {\bibinfo {volume} {22}},\ \bibinfo {pages} {9748} (\bibinfo {year} {2022})}\BibitemShut {NoStop}%
    \bibitem [{\citenamefont {Fröch}\ \emph {et~al.}(2021)\citenamefont {Fröch}, \citenamefont {Li}, \citenamefont {Chen}, \citenamefont {Toth}, \citenamefont {Kianinia}, \citenamefont {Kim},\ and\ \citenamefont {Aharonovich}}]{Froech:2021}%
      \BibitemOpen
      \bibfield  {author} {\bibinfo {author} {\bibfnamefont {J.~E.}\ \bibnamefont {Fröch}}, \bibinfo {author} {\bibfnamefont {C.}~\bibnamefont {Li}}, \bibinfo {author} {\bibfnamefont {Y.}~\bibnamefont {Chen}}, \bibinfo {author} {\bibfnamefont {M.}~\bibnamefont {Toth}}, \bibinfo {author} {\bibfnamefont {M.}~\bibnamefont {Kianinia}}, \bibinfo {author} {\bibfnamefont {S.}~\bibnamefont {Kim}},\ and\ \bibinfo {author} {\bibfnamefont {I.}~\bibnamefont {Aharonovich}},\ }\bibfield  {title} {\bibinfo {title} {Purcell enhancement of a cavity‐coupled emitter in hexagonal boron nitride},\ }\href {https://doi.org/10.1002/smll.202104805} {\bibfield  {journal} {\bibinfo  {journal} {Small}\ }\textbf {\bibinfo {volume} {18}},\ \bibinfo {pages} {2104805} (\bibinfo {year} {2021})}\BibitemShut {NoStop}%
    \bibitem [{\citenamefont {Nonahal}\ \emph {et~al.}(2023{\natexlab{a}})\citenamefont {Nonahal}, \citenamefont {Li}, \citenamefont {Ren}, \citenamefont {Spencer}, \citenamefont {Kianinia}, \citenamefont {Toth},\ and\ \citenamefont {Aharonovich}}]{Nonahal:2023}%
      \BibitemOpen
      \bibfield  {author} {\bibinfo {author} {\bibfnamefont {M.}~\bibnamefont {Nonahal}}, \bibinfo {author} {\bibfnamefont {C.}~\bibnamefont {Li}}, \bibinfo {author} {\bibfnamefont {H.}~\bibnamefont {Ren}}, \bibinfo {author} {\bibfnamefont {L.}~\bibnamefont {Spencer}}, \bibinfo {author} {\bibfnamefont {M.}~\bibnamefont {Kianinia}}, \bibinfo {author} {\bibfnamefont {M.}~\bibnamefont {Toth}},\ and\ \bibinfo {author} {\bibfnamefont {I.}~\bibnamefont {Aharonovich}},\ }\bibfield  {title} {\bibinfo {title} {Engineering quantum nanophotonic components from hexagonal boron nitride},\ }\href {https://doi.org/10.1002/lpor.202300019} {\bibfield  {journal} {\bibinfo  {journal} {Laser Photonics Rev.}\ }\textbf {\bibinfo {volume} {17}},\ \bibinfo {pages} {2300019} (\bibinfo {year} {2023}{\natexlab{a}})}\BibitemShut {NoStop}%
    \bibitem [{\citenamefont {Nonahal}\ \emph {et~al.}(2023{\natexlab{b}})\citenamefont {Nonahal}, \citenamefont {Horder}, \citenamefont {Gale}, \citenamefont {Ding}, \citenamefont {Li}, \citenamefont {Hennessey}, \citenamefont {Ha}, \citenamefont {Toth},\ and\ \citenamefont {Aharonovich}}]{Nonahal:2023a}%
      \BibitemOpen
      \bibfield  {author} {\bibinfo {author} {\bibfnamefont {M.}~\bibnamefont {Nonahal}}, \bibinfo {author} {\bibfnamefont {J.}~\bibnamefont {Horder}}, \bibinfo {author} {\bibfnamefont {A.}~\bibnamefont {Gale}}, \bibinfo {author} {\bibfnamefont {L.}~\bibnamefont {Ding}}, \bibinfo {author} {\bibfnamefont {C.}~\bibnamefont {Li}}, \bibinfo {author} {\bibfnamefont {M.}~\bibnamefont {Hennessey}}, \bibinfo {author} {\bibfnamefont {S.~T.}\ \bibnamefont {Ha}}, \bibinfo {author} {\bibfnamefont {M.}~\bibnamefont {Toth}},\ and\ \bibinfo {author} {\bibfnamefont {I.}~\bibnamefont {Aharonovich}},\ }\bibfield  {title} {\bibinfo {title} {Deterministic fabrication of a coupled cavity–emitter system in hexagonal boron nitride},\ }\href {https://doi.org/10.1021/acs.nanolett.3c01836} {\bibfield  {journal} {\bibinfo  {journal} {Nano Lett.}\ }\textbf {\bibinfo {volume} {23}},\ \bibinfo {pages} {6645} (\bibinfo {year} {2023}{\natexlab{b}})}\BibitemShut {NoStop}%
    \bibitem [{\citenamefont {Blumenthal}\ \emph {et~al.}(2018)\citenamefont {Blumenthal}, \citenamefont {Heideman}, \citenamefont {Geuzebroek}, \citenamefont {Leinse},\ and\ \citenamefont {Roeloffzen}}]{Blumenthal:2018}%
      \BibitemOpen
      \bibfield  {author} {\bibinfo {author} {\bibfnamefont {D.~J.}\ \bibnamefont {Blumenthal}}, \bibinfo {author} {\bibfnamefont {R.}~\bibnamefont {Heideman}}, \bibinfo {author} {\bibfnamefont {D.}~\bibnamefont {Geuzebroek}}, \bibinfo {author} {\bibfnamefont {A.}~\bibnamefont {Leinse}},\ and\ \bibinfo {author} {\bibfnamefont {C.}~\bibnamefont {Roeloffzen}},\ }\bibfield  {title} {\bibinfo {title} {Silicon nitride in silicon photonics},\ }\href {https://doi.org/10.1109/jproc.2018.2861576} {\bibfield  {journal} {\bibinfo  {journal} {Proc. IEEE}\ }\textbf {\bibinfo {volume} {106}},\ \bibinfo {pages} {2209} (\bibinfo {year} {2018})}\BibitemShut {NoStop}%
    \bibitem [{\citenamefont {Sharma}\ \emph {et~al.}(2020)\citenamefont {Sharma}, \citenamefont {Wang}, \citenamefont {Kaushik}, \citenamefont {Cheng}, \citenamefont {Kumar}, \citenamefont {Wei},\ and\ \citenamefont {Li}}]{Sharma:2020}%
      \BibitemOpen
      \bibfield  {author} {\bibinfo {author} {\bibfnamefont {T.}~\bibnamefont {Sharma}}, \bibinfo {author} {\bibfnamefont {J.}~\bibnamefont {Wang}}, \bibinfo {author} {\bibfnamefont {B.~K.}\ \bibnamefont {Kaushik}}, \bibinfo {author} {\bibfnamefont {Z.}~\bibnamefont {Cheng}}, \bibinfo {author} {\bibfnamefont {R.}~\bibnamefont {Kumar}}, \bibinfo {author} {\bibfnamefont {Z.}~\bibnamefont {Wei}},\ and\ \bibinfo {author} {\bibfnamefont {X.}~\bibnamefont {Li}},\ }\bibfield  {title} {\bibinfo {title} {Review of recent progress on silicon nitride-based photonic integrated circuits},\ }\href {https://doi.org/10.1109/access.2020.3032186} {\bibfield  {journal} {\bibinfo  {journal} {IEEE Access}\ }\textbf {\bibinfo {volume} {8}},\ \bibinfo {pages} {195436} (\bibinfo {year} {2020})}\BibitemShut {NoStop}%
    \bibitem [{\citenamefont {Xiang}\ \emph {et~al.}(2022)\citenamefont {Xiang}, \citenamefont {Jin},\ and\ \citenamefont {Bowers}}]{Xiang:2022}%
      \BibitemOpen
      \bibfield  {author} {\bibinfo {author} {\bibfnamefont {C.}~\bibnamefont {Xiang}}, \bibinfo {author} {\bibfnamefont {W.}~\bibnamefont {Jin}},\ and\ \bibinfo {author} {\bibfnamefont {J.~E.}\ \bibnamefont {Bowers}},\ }\bibfield  {title} {\bibinfo {title} {Silicon nitride passive and active photonic integrated circuits: trends and prospects},\ }\href {https://doi.org/10.1364/prj.452936} {\bibfield  {journal} {\bibinfo  {journal} {Photonics Research}\ }\textbf {\bibinfo {volume} {10}},\ \bibinfo {pages} {A82} (\bibinfo {year} {2022})}\BibitemShut {NoStop}%
    \bibitem [{\citenamefont {Jungwirth}\ \emph {et~al.}(2016)\citenamefont {Jungwirth}, \citenamefont {Calderon}, \citenamefont {Ji}, \citenamefont {Spencer}, \citenamefont {Flatté},\ and\ \citenamefont {Fuchs}}]{Jungwirth:2016}%
      \BibitemOpen
      \bibfield  {author} {\bibinfo {author} {\bibfnamefont {N.~R.}\ \bibnamefont {Jungwirth}}, \bibinfo {author} {\bibfnamefont {B.}~\bibnamefont {Calderon}}, \bibinfo {author} {\bibfnamefont {Y.}~\bibnamefont {Ji}}, \bibinfo {author} {\bibfnamefont {M.~G.}\ \bibnamefont {Spencer}}, \bibinfo {author} {\bibfnamefont {M.~E.}\ \bibnamefont {Flatté}},\ and\ \bibinfo {author} {\bibfnamefont {G.~D.}\ \bibnamefont {Fuchs}},\ }\bibfield  {title} {\bibinfo {title} {Temperature dependence of wavelength selectable zero-phonon emission from single defects in hexagonal boron nitride},\ }\href {https://doi.org/10.1021/acs.nanolett.6b01987} {\bibfield  {journal} {\bibinfo  {journal} {Nano Lett.}\ }\textbf {\bibinfo {volume} {16}},\ \bibinfo {pages} {6052} (\bibinfo {year} {2016})}\BibitemShut {NoStop}%
    \bibitem [{\citenamefont {Tawfik}\ \emph {et~al.}(2017)\citenamefont {Tawfik}, \citenamefont {Ali}, \citenamefont {Fronzi}, \citenamefont {Kianinia}, \citenamefont {Tran}, \citenamefont {Stampfl}, \citenamefont {Aharonovich}, \citenamefont {Toth},\ and\ \citenamefont {Ford}}]{Tawfik:2017}%
      \BibitemOpen
      \bibfield  {author} {\bibinfo {author} {\bibfnamefont {S.~A.}\ \bibnamefont {Tawfik}}, \bibinfo {author} {\bibfnamefont {S.}~\bibnamefont {Ali}}, \bibinfo {author} {\bibfnamefont {M.}~\bibnamefont {Fronzi}}, \bibinfo {author} {\bibfnamefont {M.}~\bibnamefont {Kianinia}}, \bibinfo {author} {\bibfnamefont {T.~T.}\ \bibnamefont {Tran}}, \bibinfo {author} {\bibfnamefont {C.}~\bibnamefont {Stampfl}}, \bibinfo {author} {\bibfnamefont {I.}~\bibnamefont {Aharonovich}}, \bibinfo {author} {\bibfnamefont {M.}~\bibnamefont {Toth}},\ and\ \bibinfo {author} {\bibfnamefont {M.~J.}\ \bibnamefont {Ford}},\ }\bibfield  {title} {\bibinfo {title} {First-principles investigation of quantum emission from hbn defects},\ }\href {https://doi.org/10.1039/c7nr04270a} {\bibfield  {journal} {\bibinfo  {journal} {Nanoscale}\ }\textbf {\bibinfo {volume} {9}},\ \bibinfo {pages} {13575} (\bibinfo {year} {2017})}\BibitemShut {NoStop}%
    \end{thebibliography}
\end{document}